\documentclass[useAMS,usenatbib]{mnras}

\usepackage{psfrag,amsmath,amssymb,color,epsfig,rotating}
\usepackage{graphicx}
\usepackage{multirow}
\usepackage{tabularx}
\usepackage{geometry}
\usepackage{array}

\usepackage[dvipsnames]{xcolor}
\def\HI{{\rm HI}\,} 
\def\k{{\mathbf{k}}} 
 
\def\k{{\mathbf{k}}}
 
\def\k{{\bf k}}
\def\kpe{k_{\perp}}
\def\kpa{k_{\parallel}}

\definecolor{jade}{rgb}{0.0, 0.66, 0.42}

\definecolor{deepcrimson}{rgb}{0.8, 0.0, 0.15}

\begin{document}

\title [RSD in the post-reionization 21-cm power
  spectrum]{Modelling  redshift-space distortion in the
  post-reionization \HI 21-cm power spectrum}  

\author[Sarkar et al.]{Debanjan
  Sarkar$^{1}$\thanks{debanjan@cts.iitkgp.ernet.in},Somnath Bharadwaj$^{1,
    2}$\thanks{somnath@phy.iitkgp.ernet.in}\\ $^1$Centre for Theoretical Studies, Indian
  Institute of Technology Kharagpur, Kharagpur - 721302, India \\ $^2$Department of Physics, Indian Institute of Technology Kharagpur,
  Kharagpur - 721302, India}

\date{} \maketitle

\begin{abstract}
The post-reionization \HI 21-cm signal is an excellent candidate for
precision cosmology, this however requires accurate modelling of the
expected signal.  \citet{sarkar-bharadwaj-2016} have simulated the
real space \HI 21-cm signal, and have modelled the \HI power spectrum
as $P_{\HI}(k)=b^2 P(k)$ where $P(k)$ is the dark matter power
spectrum and $b(k)$ is a (possibly complex) scale dependent bias for
which fitting formulas have been provided. This paper extends these
simulations to incorporate redshift space distortion and predict the
expected redshift space \HI 21-cm power spectrum
$P^s_{\HI}(\kpe,\kpa)$ using two different prescriptions for the
\HI distributions and peculiar velocities. We model
$P^s_{\HI}(\kpe,\kpa)$ assuming that it is the product of
$P_{\HI}(k)=b^2 P(k)$ with a Kaiser enhancement term and a Finger of
God (FoG) damping which has $\sigma_p$ the pair velocity dispersion as
a free parameter.  Considering several possibilities for the bias and
the damping profile, we find that the models with a scale dependent
bias and a Lorentzian damping profile best fit the simulated
$P^s_{\HI}(\kpe,\kpa)$ over the entire range $1 \le z \le 6$.  The
best fit value of $\sigma_p$ falls approximately as  $(1+z)^{-m}$
with $m=2$ and $1.2$ respectively for the two different
prescriptions.  The model predictions are consistent with the
simulations for $k < 0.3 \, {\rm Mpc}^{-1}$ over the entire $z$ range
for the monopole $P^s_0(k)$, and at $z \le 3$ for the quadrupole
$P^s_2(k)$.  At $z \ge 4$ the models underpredict $P^s_2(k)$ at large
$k$, and the fit is restricted to $k < 0.15 \, {\rm Mpc}^{-1}$.
\end{abstract}


\section{Introduction}
\label{sec:introduction}

After the Cosmic Microwave Background (CMB) radiation, the
cosmological 21-cm background radiation is one of the most interesting
observational frontiers.  This originates from the spin-flip
transition in the ground state of neutral hydrogen (\HI).  The
cumulative redshifted 21-cm emission from all the \HI sources forms a
diffused background radiation.  A statistical detection of the
intensity fluctuations in this 21-cm background provides us a unique
way of quantifying the source clustering
\citep{bharadwaj-nath-sethi01,bharadwaj-sethi01}.  This technique,
widely known as 21-cm intensity mapping, provides a three dimensional
view of large-scale structures in the Universe and makes it possible to
survey large volumes of space using currently functioning and upcoming
radio telescopes
\citep{bharadwaj-pandey03,bharadwaj-ali05,wyithe-loeb08-fluctuations-in-21cm}.
In the post-reionization era ($z \le 6$), the 21-cm signal is mostly
unaffected by the complex reionization processes and the 21-cm power
spectrum is proportional to the underlying matter power spectrum
\citep{wyithe-loeb09}.  A detection of the Baryon Acoustic
Oscillations (BAO) in the 21-cm power spectrum can place tight
constraints on the equation of state of dark energy
\citep{wyithe-loeb-geil08,chang-pen-peterson08,seo-dodelson10,masui-mcdonald-pen10}. An
accurate measurement of the 21-cm power spectrum can also provide
independent estimates of the different cosmological parameters
\citep{loeb-wyithe08,bharadwaj-sethi-saini09} without reference to the
BAO.  \citet{pourtsidou15-testing-gravity-HI-IM} have investigated the
possibility of testing Einstein's general theory of relativity (GR)
and the standard cosmological model via the $E_G$ statistic
\citep{zhang-ligouri07-probing-gravity-EG-statistics} using 21-cm
intensity mapping.

Several 21-cm intensity mapping experiments like BAOBAB
\citep{pober-parsons13}, BINGO \citep{battye-brown12}, CHIME
\citep{bandura14}, the Tianlai project
\citep{chen12-tianlai,chen-wang-16-tianlai}, GBT-HIM
\citep{chang16-GBT-HIM}, SKA1-MID/SUR
\citep{bull-ferreira15-late-time-cosmology-with-21cm-IM} have been
planned to cover the intermediate-redshift range $z \sim\, 0.5 \,-\,
2.5$.  Their primary goal is to measure the comoving scale of BAO
around the onset of acceleration at $z \sim\,1$.  Efforts are also
underway to observe the 21-cm intensity fluctuations around $z \sim\,
3.35$ using OWFA \citep{subrahmanya-manoharan-chengalur17-OWFA}.  On
the other hand, the upgraded GMRT \citep[uGMRT;][]{gupta17-uGMRT} and
the upcoming SKA2 \citep{santos-bull15-SKA-HI-IM-survey} promises to
cover a large redshift range.

The 21-cm signal is intrinsically very weak, and it 
is very important to correctly model the expected  signal
in order
to make realistic predictions for the various upcoming experiments
\citep[{\textit e.g.}][]{bull-ferreira15-late-time-cosmology-with-21cm-IM,sarkar-bharadwaj-ali17-OWFA}. 
Modelling is also  important in order  
to correctly interpret the 21-cm signal once it is detected.

There has been considerable work to model the post-reionization 
\HI distribution  and the expected 21-cm signal using 
semi-numerical  prescriptions  coupled
with N-body simulations
\citep{bagla10, khandai-sethi-dimatteo11,
  tapomoy-mitra-majumder12, seehars-paranjape15}. In subsequent
works,  semi-numerical  prescriptions have also been coupled
with  hydrodynamic simulations \citep{dave-katz-openheimer13,
  villaescusa-navarro-viel-datta-choudhury14, kim-wyithe-baugh-lagos16, villaescusa-navarro16-HI-in-galaxy-clusters}.
A number of analytical frameworks have also been developed 
\citep{marin-gnedin10,hamsa-refregier-amara16-halomodel2,castorina-villaescusa-navarro16,aurelie-umeh-santos17-HIbias}.

In a recent  work (\citealt{sarkar-bharadwaj-2016}, hereafter Paper I),
we have used a semi-numerical technique coupled with N-body
simulations to model the \HI distribution in the redshift range
$1\le z \le 6$. We have quantified the evolution
of the \HI bias across this $z$ range for $k$ values in the
range $0.04 \le k/{\rm Mpc}^{-1} \le 10$, and we provide
polynomial fitting formulas for the bias across this 
$k$ and $z$ range. Paper I and most of  the other works 
discussed above  are however limited by the
fact that they have modelled  the real space \HI distribution
ignoring the redshift space distortion (RSD)  introduced
by  peculiar velocities. It is important to note that
RSD plays an important role in the 21-cm intensity mapping signal
\citep{bharadwaj-ali04}. In this paper we have extended the work
presented in Paper I to include peculiar velocities and 
use this to model the RSD in the 21-cm power spectrum.

The RSD in galaxy redshift surveys has proven to be an
important field of study (see \citealt{hamilton-98-rsd-review} for a
review).  It offers a distinct way of estimating the cosmological 
parameters and/or any departure from the standard theory of structure
formation
\citep{cole-percival-peacock05-cosmology-2df,parkinson12-cosmological-results-WiggleZ-final-data-release,song-sabiu14-cosmological-tests-BOSS-DR11,
  li-park16-cosmological-constraints-BOSS-DR12}.
This also gives an independent estimate of the cosmological expansion
rate
\citep{guzzo-pierleoni08-nature-of-cosmic-acceleration-using-galaxy-redshift-distortions,
  kazin-tinker12-testing-gravity-and-cosmic-acceleration-with-galaxy-clustering} 
and can also help to break the degeneracy between various models of
modified gravity
\citep{johnson-blake-16-modified-gravity-using-galaxy-peculiar-velocities,
  de_la_torre16-gravity-test-from-RSD-and-lensing}.  We expect to get
similar information from the study of RSD effects in 21-cm intensity
mapping experiments \citep{raccanelli-bull-camera15-RSD-with-SKA,
  santos-bull15-SKA-HI-IM-survey,bull-ferreira15-late-time-cosmology-with-21cm-IM}.

In Paper I we have used N-body simulations and a semi-numerical
technique, originally proposed by \citet{bagla10}, to simulate the real
space \HI distribution and model the real space \HI power spectrum
$P_{\HI}(k)$  over the redshift range $1 \le z \le 6$. In the present
work we have used the same simulations  with the addition that we 
have used the peculiar velocities from the N-body simulations to map
the \HI distribution to redshift space. In this work we model the RSD
effect in the redshift space \HI power spectrum $P^{s}_{\HI}
(\kpe,\kpa)$, and  use this to study  its various  moments - namely
the monopole $P^s_0(k)$ and the quadrupole $P^s_2(k)$.

A brief outline of the paper follows. We present the redshift space
\HI simulations in Section~\ref{sec:simulating-RSD-HI}, and the
simulated redshift space \HI power spectrum $P^{s}_{\HI}(\kpe,\kpa)$
is presented in Sub-Section~\ref{sub-sec:HIrsdpower}. In
Section~\ref{sec:model-power-spec} we outline the different models
which we have considered for $P^{s}_{\HI}(\kpe,\kpa)$, the $\chi^2$
minimization used to fit the models to the simulations is also
discussed here.  We present the Results in Section~\ref{sec:results},
and Section~\ref{sec:summary} presents Summary and Discussion.

We have adopted the  best 
fit cosmological parameters from \citet{planck-collaboration15}.

\section{Simulating  \HI distribution in redshift-space}
\label{sec:simulating-RSD-HI}

We first simulate the post-reionization \HI distribution in real space.
Here we follow the methodology which has been extensively
described in Paper I.  We use a gravity only Particle Mesh N-body code
\citep{bharadwaj-srikant04} to generate snapshots of the dark matter
distribution at the desired redshifts.  The simulations contain
$[1,072]^3$ dark matter particles in a $[2,144]^3$ regular cubic
grid of spacing $0.07~{\rm Mpc}$ with a total simulation volume
(comoving) of $[150.08~{\rm Mpc}]^3$.  The grid spacing limits the
mass resolution of our simulations to $10^8\, {\rm M_{\odot}}$. Our
simulations span the redshift range $z=1$ to $6$ with redshift
interval $\Delta z=0.5\,$.

We use the ``Friends-of-Friends'' (FoF), \citep{davis85} algorithm to
locate the haloes in the dark matter distribution. We have used a
linking length $l_f=0.2$ (in the unit of mean inter particle
separation) and set a criterion that a halo must contain at least $10$
dark matter particles and hence we can resolve haloes with mass
$10^9\, {\rm M_{\odot}}$ or larger in our simulations.  Our halo mass
resolution is consistent with a recent study
\citep{kim-wyithe-baugh-lagos16} which shows that for $z \, \geq 0.5$,
we require a dark matter halo resolution better than $\sim\,
10^{10}\,h^{-1} {\rm M_{\odot}}$ to ensure that the predicted 21-cm
brightness temperature fluctuations are well converged.

The Inter Galactic Medium (IGM) is highly ionized in the
post-reionization era \citep{fan-carilli-keating06,fan-strauss06}.
The bulk of the \HI at these redshifts is believed to reside in
self-shielded gas clouds that appear as Damped Lyman-$\alpha$ systems
(DLAs) in quasar spectra
\citep{storrie-lomb00,prochaska-herbertfort-wolfe05,zafar-peroux-popping-milliard13}.
These clouds are further believed to be associated with galaxies
\citep{haehnelt-steinmetz-rauch00} which are hosted in dark matter
haloes.  \HI therefore is a tracer of the galaxy distribution.
\citet{bouche-gardner05-DLA-Halo-mass} measured the clustering of DLAs
at $z \sim 3$ in a numerical simulation and found that DLAs occupy
moderate mass haloes with an upper limit of $\log(\, M_h/M_{\odot})
\sim 11.1$.  Again a cross correlation study between DLAs and Lyman
Break Galaxies (LBGs) at $z \sim 3$ suggests that DLAs preferably
reside in haloes having mass $M_h$ in the range $10^9 < \,
M_h/M_{\odot} \, < 10^{11}$ \citep{cooke-wolfe-gawiser06} which is in
agreement with the findings of \cite{pontzen08} based on numerical
simulations.  However, in a similar work with large volume
simulations, \citet{cen12-nature-of-DLAs-and-their-hosts} found that
DLAs prefer to reside in relatively more massive haloes ($10^{10} < \,
M_h/M_{\odot} \, < 10^{12}$ at $1.6 < \,z\, <4$) where galactic winds
control their kinematics.  A cross-correlation analysis of DLAs and
the Lyman$-\alpha$ forest at $z\sim 3$
\citep{font-ribera12-cross-correlation-DLA-Ly-alpha} suggests that the
DLAs favour relatively massive haloes with mass $M_h = 6 \times
10^{11} M_{\odot}$. Note that this mass limit has been calculated by 
assuming a relation between the DLAs cross-section and the host halo mass, 
and DLAs can be hosted by the smaller mass haloes if the slope of this 
relation is steep enough.

In this work we assume that the \HI resides solely in the dark matter
haloes, and the \HI mass $M_{\HI}$ in a halo depends only on the halo
mass independent of the environment of the halo.  We expect $M_{\HI}$
to increase with $M_h$, however, observations at very low redshift
suggests that the massive elliptical galaxies or galaxy clusters
contain very little \HI \citep[eg. see][ and references
therein]{serra-oosterloo-morganti12}.  This indicates that haloes with
mass greater than a maximum value, $M_{max}$, will contain little or
no \HI.  Further, we do not expect haloes with masses below a lower
limit $M_{min}$ to contain \HI as they would not be able to shield the
neutral gas from the harsh ionizing background radiation.  Based on
these arguments, \citet{bagla10} have proposed three schemes for
populating the simulated haloes with \HI.  In Paper I and our current
work we have used the third scheme of \citet{bagla10} to populate the
haloes.  This scheme considers a redshift dependent relation between
the $M_h$ and circular velocity $v_{circ}$
%----------------------------------
\begin{equation} M_h \simeq 10^{10} \,{\rm M}_{\odot}\, \Big( \frac{v_{circ}}{60 \, {\rm
km}\, {\rm s}^{-1}} \Big)^3 \Big( \frac{1+z}{4} \Big)^{-\frac{3}{2}}\,.
\label{eq:virial_relation}
\end{equation} 
%---------------------------------- 
It is assumed that only haloes with a minimum circular velocity
$v_{circ} = 30 \, {\rm km}\, {\rm s}^{-1}$ will host \HI, which sets the lower
mass limit $M_{min}$ and $v_{circ} = 200 \, {\rm km}\, {\rm s}^{-1}$ defines
the upper mass limit $M_{max}$.

The \HI mass in a halo is related to the halo mass as 
%----------------------------------
\begin{equation}
  M_{\rm \HI}(M_h) = \left\{  \begin{array}{l l}	 
  f_3\frac{M_h}{1+\left(\frac{M_h}{M_{\rm max}}\right)} & \quad
  \text{if $M_{\rm min}\leqslant M_h$} \\ 0 & \quad
  \text{otherwise}\\
  \end{array} \right.  \,.
  \label{eq:bagla_scheme}
\end{equation}
%----------------------------------
where $f_3$ is a free parameter which controls the total \HI content
in our simulations. The choice of $f_3$ does not affect the results of
this work, and we have used $f_3$ such that $\Omega_{\HI}$ in our
simulations remains fixed at a value $\sim\, 10^{-3}$.
Here we have placed
all the \HI at the halo center of mass. This is a reasonably good assumption
in real space as the intensity mapping experiments  do not have 
adequate  angular  resolution to resolve the \HI distribution within
individual  galaxies at high  redshifts.  We refer to this method as
`HC' and  Paper I is entirely based on  this  method.

Our simulations have a fixed halo mass resolution of 
$M_h=10^9 \,{\rm M}_{\odot}$ and at $z>3.5$, $M_{min}$ falls below
 this mass resolution. We have used $M_{min}=10^9 \,{\rm M}_{\odot}$ 
for $z>3.5$ in the \HI assignment scheme (eq.~\ref{eq:bagla_scheme}). 
For $z \le 3.5$, $M_{min}$ is above our halo mass resolution and we 
can fully resolve the smallest haloes where \HI resides. 
The effect of choosing $M_{min}=10^9 \,{\rm M}_{\odot}$ for $z>3.5$ 
has been discussed in Paper I.

The steps described till now yield the real space \HI distribution
which was analyzed in Paper I.  We now map this \HI distribution to
redshift space using the line of sight component of the peculiar
velocity ${\bf v}$.  For a distant observer along the $z-$axis, the
halo position in redshift space ${\bf s}$ is related to its real
position as
%----------------------------------
\begin{equation} 
{\bf s}={\bf x} \,+\, \frac{ ({\bf v} . \hat{z}) \hat{z}}{a\, H(a)} \,,
\label{eq:redshift_mapping}
\end{equation} 
%---------------------------------- 
where, $a$ and $H(a)$ are the scale factor and Hubble parameter
respectively.

In the HC method we have assumed that the \HI in a
halo  moves with the mean velocity of the host halo. In this method,
in addition to placing the entire \HI at the halo center of mass, we
have also assigned the mean velocity of the host halo to the
\HI. This method ignores the motion of the \HI  within the halo.
While the intensity mapping experiments  lack the angular resolution to
resolve the \HI distribution within individual galaxies, they do have
the frequency resolution required to make out the motion of the \HI
within  individual haloes.  Further, the effect of these motions
may also extend to large length-scales through the Finger of God (FoG)
effect \citep{jackson72-FoG}. It is therefore desirable to include
the velocity dispersion of the \HI within the individual  haloes. 
We presently have very little or no knowledge of how the \HI is distributed
within the haloes at high redshifts. Low redshift observations suggest
that the \HI is largely contained in the disks of   spiral galaxies.
 On the other hand, there can be more than one galaxy in a halo.
A detailed modelling of  the galactic \HI disks is  beyond
the scope of  this paper, and we have not attempted this here.  In the
present work we have used a simple method to account for the \HI
velocity dispersion   within the individual haloes. We have uniformly
distributed the \HI content of a halo among all the 
particles that constitute the halo. Further, the \HI is assumed to
move with the same velocity as the corresponding  particle. We refer
to this as the halo-particle `HP' method.  While neither the HC method
nor the HP method is expected to give  a very realistic  picture of
the \HI distribution, it would still be quite reasonable to 
treat the two methods as respectively providing the lower and upper
limits  to the velocity dispersion effects.  Both these methods are
however expected to incorporate  the coherent motions of the \HI
reasonably well.

Before proceeding to the redshift space power spectrum, it is
important to note that the real space \HI distribution and
consequently the real space \HI power spectrum differs between HC and
HP methods. This difference is minimum at $z=3$ where the real space
power spectra differ by $\lesssim 1 \%$ at $k \le 1 \,{\rm Mpc}^{-1}$
and $\lesssim 10 \%$ at $k \le 3 \,{\rm Mpc}^{-1}$. The difference
increases at higher ($z \geq 4$) and lower ($z=2$) redshifts where it
hovers between $\lesssim 3 - 6 \%$ at $k \le 1 \,{\rm Mpc}^{-1}$ and
$\lesssim 10 \%$ at $k \le 3 \,{\rm Mpc}^{-1}$. This difference is
maximum at $z=1$ where the real space power spectra differ by
$\lesssim 10 \%$ at $k \le 0.4 \,{\rm Mpc}^{-1}$ and $\lesssim 15 \%$
at $k \le 1 \,{\rm Mpc}^{-1}$.  These differences are relatively small
given our current lack of understanding about how the \HI is
distributed at these redshifts. Throughout the present work, for the
real space power spectra, we have ignored these small differences
between the HC and HP methods and have used the results from Paper I
to model the values of the \HI bias parameters.

We have considered five statistically independent
realizations of the simulation to estimate the mean and variance for
all the results presented here.

%-----------------------------------------------------
\begin{figure*}
	
	\psfrag{z=1.0}[c][c][1.5][0]{$z=1\quad$}
	\psfrag{z=2.0}[c][c][1.5][0]{$z=2\quad$}
	\psfrag{z=3.0}[c][c][1.5][0]{$z=3\quad$}
	\psfrag{z=4.0}[c][c][1.5][0]{$z=4\quad$}
	\psfrag{z=5.0}[c][c][1.5][0]{$z=5\quad$}
	\psfrag{z=6.0}[c][c][1.5][0]{$z=6\quad$}
	\psfrag{kperp}[c][c][1.5][0]{$\kpe~{\rm Mpc}^{-1}$}
	\psfrag{kpar}[c][c][1.5][0]{$\kpa~{\rm Mpc}^{-1}$}
	
	\centering
	\includegraphics[width=0.6\textwidth,angle=-90]{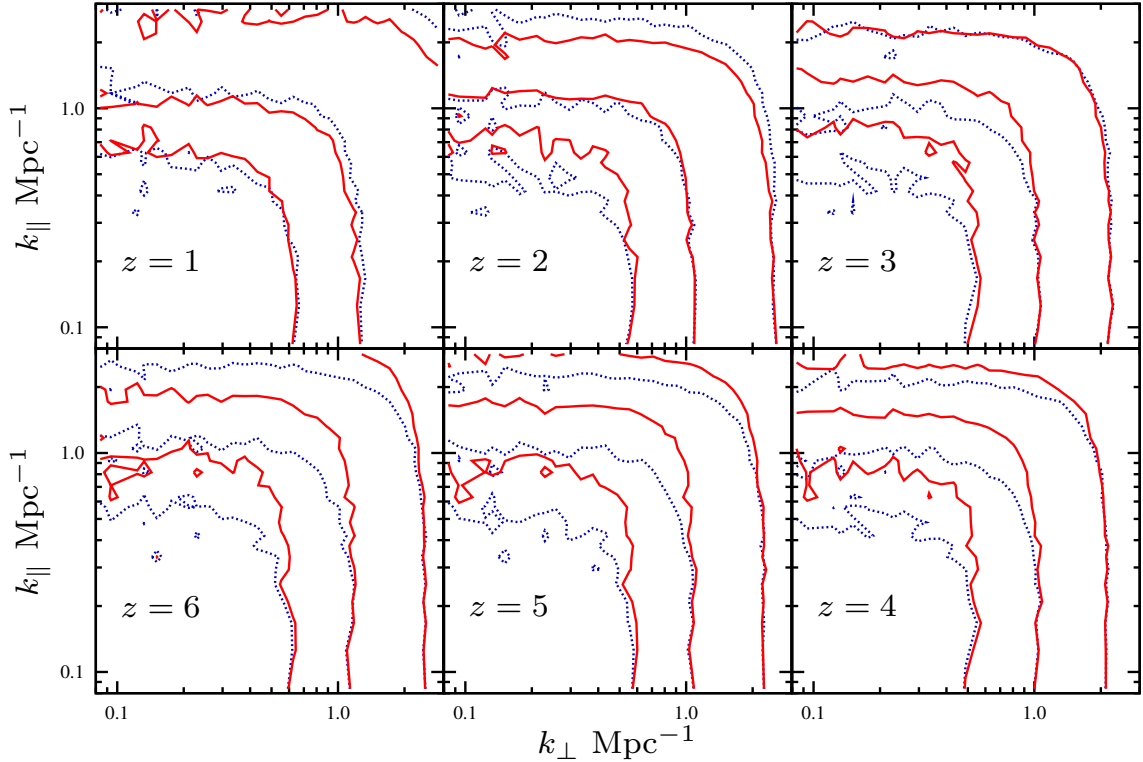}

	\caption{The red solid contours show the simulated redshift
          space \HI power spectrum $P^s_{\HI}(\kpe,\kpa)$ at six different
          redshifts, while the blue dotted contours show the real space
          counterpart. Both the power spectra are calculated using the HC method.The contour values increase inwards. }  
	\label{fig:HI_power_spectrum}
\end{figure*}
%-------------------------------------------------------------

%-----------------------------------------------------
\begin{figure*}
	
	\psfrag{z=1.0}[c][c][1.5][0]{$z=1\quad$}
	\psfrag{z=2.0}[c][c][1.5][0]{$z=2\quad$}
	\psfrag{z=3.0}[c][c][1.5][0]{$z=3\quad$}
	\psfrag{z=4.0}[c][c][1.5][0]{$z=4\quad$}
	\psfrag{z=5.0}[c][c][1.5][0]{$z=5\quad$}
	\psfrag{z=6.0}[c][c][1.5][0]{$z=6\quad$}
	\psfrag{kperp}[c][c][1.5][0]{$\kpe~{\rm Mpc}^{-1}$}
	\psfrag{kpar}[c][c][1.5][0]{$\kpa~{\rm Mpc}^{-1}$}
	
	\centering
	\includegraphics[width=0.6\textwidth,angle=-90]{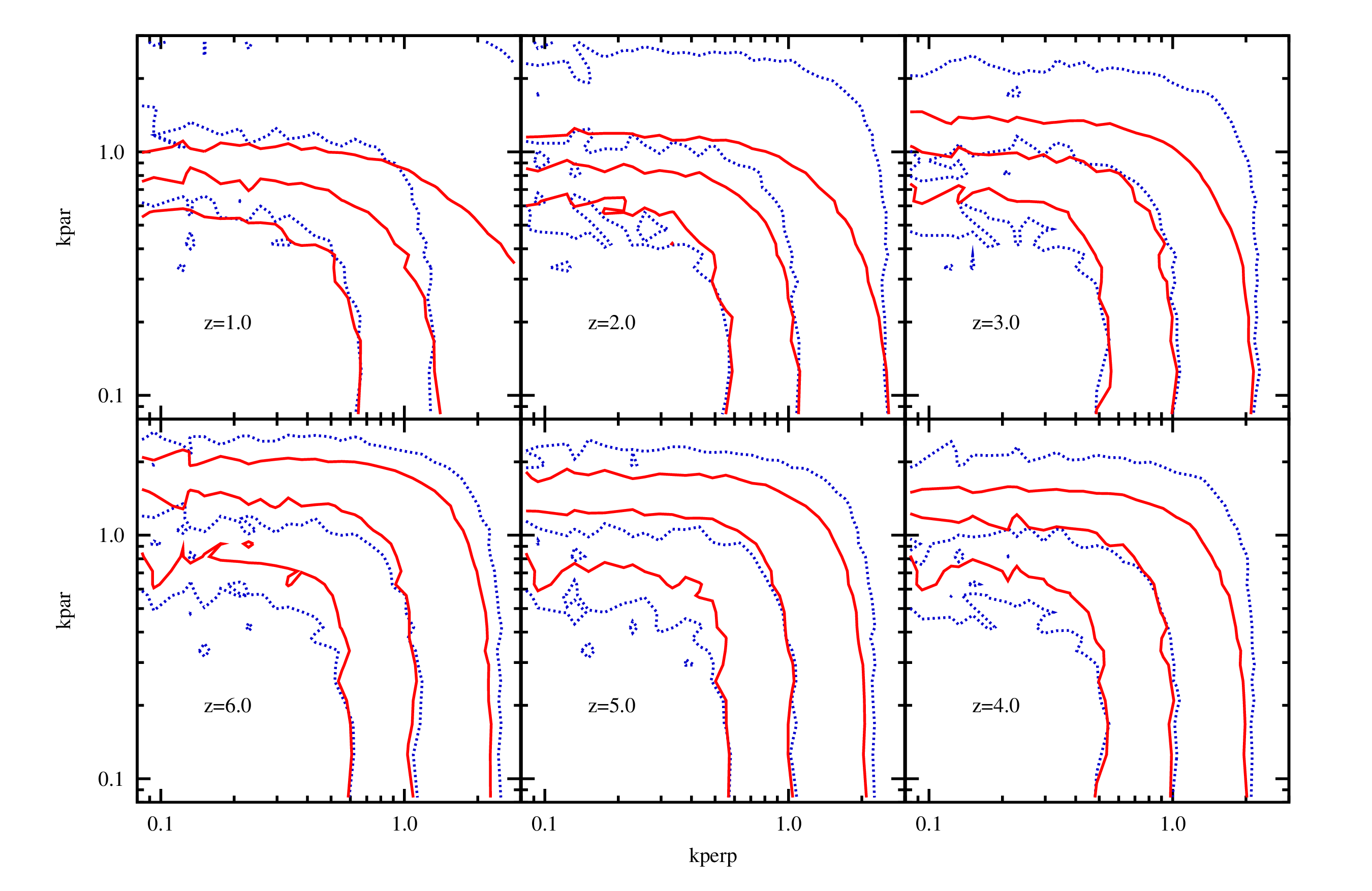}

	\caption{The red solid contours show the simulated redshift
		space \HI power spectrum  $P^s_{\HI}(\kpe,\kpa)$ for the HP method at six
		different redshifts, while  the  blue dotted contours show
		the real space counterpart.  The contour values increase
		inwards. }  
	\label{fig:HI_power_spectrum_HP}
\end{figure*}
%-------------------------------------------------------------

\subsection{The redshift space \HI power spectrum}
\label{sub-sec:HIrsdpower}

The real space \HI power spectrum $P_{\HI}(\k)$, studied in Paper I,
is isotropic $P_{\HI}(\k)=P_{\HI}(k)$.  The peculiar velocity
introduces an anisotropy along the line of sight direction.  We
quantify the \HI distribution in redshift space in terms of the power
spectrum, $P^s_{\HI}(k,\mu) \equiv P^s_{\HI}(\kpe,\kpa)$.  Here,
$\kpa$ is the line of sight component of the wave vector $\k$,
$\k_{\perp}$ is the component, perpendicular to the line of sight and
$\kpe=\mid \k_{\perp} \mid$, and $\mu=\kpa/k$.  The peculiar velocity
has broadly two effects.  The first is an enhancement of power on
large scales owing to the streaming of matter into over-dense regions
and out of under-dense regions\cite[Kaiser effect;][]{kaiser87}.  The
second effect is a reduction of power on small scales due to the
random motions \citep[Finger-of-God (FoG) effect;][]{jackson72-FoG}.
Figure~\ref{fig:HI_power_spectrum} shows the simulated \HI power
spectrum $P_{\HI}(\kpe,\kpa)$ using the HC method in both real (blue
dotted contours) and redshift space (red solid contours) at six
distinct redshifts.  Here, the outer contours correspond to smaller
spatial scales (larger $k$), and the inner contours correspond to
larger spatial scales (smaller $k$).  The contour values increase
inwards.  The real space \HI power spectrum $P_{\HI}(k)$, which is
isotropic, has been shown for comparison.  We see that the
$P^s_{\HI}(\kpe,\kpa)$ contours are elongated along the $\kpa$
direction for smaller $k$ values and compressed for larger $k$ values.
This elongation is due to the Kaiser effect and the compression is
caused by the FoG.  Now, for $z \geq 4$, the FoG damping is not
present in the power spectrum for the $k$ range that we have chosen
for plotting.  This means that for the HC method the effect of
the small scale FoG damping is not very large in these redshifts.
While the FoG compression is visible at lower redshifts ($z\le 2$)
this is not very strong because the HC method does not include the
velocity dispersion within the
haloes. Figure~\ref{fig:HI_power_spectrum_HP} shows
$P_{\HI}(\kpe,\kpa)$ for the HP method. It can be seen that the
$P^s_{\HI}(\kpe,\kpa)$ contours look different from those of the HC
method (Figure~\ref{fig:HI_power_spectrum}). For the HP method, the
FoG compression is significantly more pronounced in comparison to the
HC method and the compression is also visible at all redshifts. At
high $z$, the FoG compression is restricted to the largest $k$ values,
while the Kaiser elongation is visible at smaller $k$. As $z$
decreases, the magnitude of the compression increases and it also
extends to smaller $k$ values. The contours of $P_{\HI}(k)$ and
$P^s_{\HI}(\kpe,\kpa)$ coincide along $\kpe$ direction for $\kpa=0$,
suggesting that the anisotropy is only along the $\kpa$ direction and
maximum along the line of sight ($\kpe=0$).  We model this anisotropy
using a dispersion model and try to explain the different effects in
terms of modelling in the next section.

\section{Modelling the redshift space \HI power spectrum and its moments}
\label{sec:model-power-spec}

The simplest model for redshift space distortion
%--------------------------------------------
\begin{equation} P_{\HI}^s(\kpe,\kpa)=b^2 \left(1 + \beta \mu^2 \right)^2
P(k) D_{FoG}(\kpa,\sigma_p)\,.
\label{eq:kaiser_model}
\end{equation}
%--------------------------------------------
combines the Kaiser enhancement $\left(1 + \beta \mu^2 \right)^2$ with
an independent small-scale suppression $D_{FoG}(\kpa,\sigma_p)$ to
account for the FoG effect. Such models have been extensively used
\citep{peacock92,park-vogeley94,peacock-dodds94,
ballinger-peacock-heavens96} for the power spectrum of a variety of
other tracers such as galaxies.  Here $\beta= f/b$ is the redshift
distortion parameter with $f$ being the logarithmic growth rate and
$b$ the linear scale-independent bias, $\mu=\kpa/k$ is the cosine of
the angle between $\k$ and the line of sight, $\sigma_p$ is the
pairwise velocity dispersion, and $P(k)$ is the real space dark matter
power spectrum.  Here, $\sigma_p$ is in units of comoving Mpc, we can
equivalently use $[\sigma_p a H(a)]$ in units of km/s.  There are
several forms of $D_{FoG}(\kpa,\sigma_p)$ which have been used,
however the Lorentzian and the Gaussian are the most widely used
profiles.  The assumption of an exponential pairwise velocity
distribution with a scale independent width leads to a Lorentzian
profile in Fourier space \citep{davis-peebles83,
hamilton-98-rsd-review, hatton-cole99, white01-RSD-halo-model,
seljak01-RSD-halo-model}. A motivation for the Gaussian damping profile
can be found in \citet{bharadwaj01-nonlinear-RSD}. The square of a
Lorentzian profile has also been used \citep{cole-fisher-weinberg95}.
Models which incorporate the Kaiser enhancement along with a
small-scale damping as above are known as ``dispersion models''.

In Paper I, we have modelled the \HI distribution through a linear,
complex, scale dependent bias $\tilde{b}(k)$.  The real space \HI
power spectrum $P_{\HI}(k)$ is related to the dark matter power
spectrum through $P_{\HI}(k)=b^2(k) P(k)$, where $b(k)= \mid
\tilde{b}(k) \mid $ is the modulus of the complex bias.  The real part
of the complex bias $b_r(k)=Re[\tilde{b}(k)]$ quantifies the
cross-correlation between the \HI and the dark matter density fields,
and $P_c(k)$, the \HI-dark matter cross-correlation power spectrum can
be expressed as $P_c(k)=b_r(k)P(k)$.  The cross-correlation can be
equivalently expressed in terms of the stochasticity parameter
$r=b_r/b$.  In this paper we use the scale dependent bias $b(k)$ and
the stochasticity parameter $r(k)$ to quantify the clustering of the
\HI relative to dark matter fluctuations.

Paper I presents a detailed study of how $b(k)$ and $b_r(k)$ vary
across the range $z \le 6$ and $0.04 \le k/{\rm Mpc}^{-1} \le 10$.
The bias is found to be scale independent at large scales (small $k$)
and exhibit scale dependence at small scales (large $k$).  The value
of the bias is found to decrease with decreasing $z$.  The
stochasticity $r(k)$ is found to be close to unity at all scales for
$z>2$.  At lower redshifts ($z \le 2$), $r \sim 1$ at large scales,
however $r<1$ at small scales.  Paper I also presents polynomial
fitting formulas in $k$ and $z$ for both $b(k)$ and $b_r(k)$.  These
formulas can be used to calculate $b(k)$ and $r(k)$ across the entire
$k$ and $z$ range considered in the present paper.  As discussed
in Section \ref{sec:simulating-RSD-HI}, there is a small difference in
$P_{\HI}(k)$ (also $b(k)$ and $r(k)$) between the HC and the HP
methods. In the present work we have ignored this and used the fitting
formulas for the \HI bias from Paper I for the rest of our analysis.

 The presence of a complex bias $\tilde{b}$, or equivalently the
stochasticity $r$, modifies the Kaiser term to ($1+ 2 r \beta \mu^2 +
\beta^2 \mu^4$) (see Appendix~\ref{sec:Appendix-A} for a  derivation).
 In this paper we have used
%--------------------------------------------
\begin{equation} P^s_{\HI}(\kpe,\kpa)=b^2 \left(1+ 2 \beta r \mu^2 +
\beta^2 \mu^4\right) P(k) D_{FoG}(\kpa,\sigma_p) \,,
\label{eq:modified_kaiser_model}
\end{equation}
 %--------------------------------------
where $P(k)$ is the real space dark matter power spectrum  from the
simulations in Paper I, and  $b$, $\beta$ and $r$ are three scale
dependent parameters whose values change with redshift (Paper I). 
Here we have considered three different cases, namely: 

(A) scale dependent bias and stochasticity as determined in Paper I. 
This essentially treats the bias as a complex quantity.

(B) scale dependent bias as determined in Paper I with $r=1$.
This is equivalent to a real bias $b= \mid \tilde{b} \mid$.

(C) scale independent bias $b=b_0$ with $r=1$.
Here the bias only evolves with $z$ but has no $k$ dependence, and 
$b_0$ is the constant terms in the polynomial which quantifies the $k$
dependence in Paper I. The value of $b_0$ essentially corresponds to
 large length-scales where we have a nearly scale
independent bias (Paper I). 

Here we have also considered three different damping profiles, namely: 

(1)  Lorentzian, $D_{FoG}(\kpa,\sigma_p)=(1+\frac{1}{2} \kpa^2 \sigma^2_p)^{-1}$

(2) Gaussian,  $D_{FoG}(\kpa,\sigma_p)=\exp{(-\frac{1}{2} \kpa^2 \sigma^2_p)}$

(3) Lorentzian squared, $D_{FoG}(\kpa,\sigma_p)=(1+\frac{1}{2} \kpa^2
\sigma^2_p)^{-2}$.

We therefore have nine possible combinations A1, A2,..., C2 and C3
which we have considered in this paper.  All the different models have
only one free parameter $\sigma_p$.  Note that, $\sigma_p$ here is in
units of comoving Mpc and we can equivalently use $[\sigma_p a H(a)]$
in units of km/s.  In order to determine how well these models are
able to capture the anisotropy in the simulated $P^s_{\HI}(\kpe,
\kpa)$ (using both the HC and HP methods), for each model we fit 
the simulated $P^s_{\HI}(\kpe, \kpa)$ to
determine the best fit value of the parameter $\sigma_p$.  The
corresponding goodness of fit for all the models is provided in the
Table~\ref{tab:chisq}.  While nearly all the models work well at small
$k$, they all exhibit very significant deviations from the simulations
at large $k$. Further, these deviations are also seen to increase at
lower redshifts. These deviations at large $k$ severely influence the
fitting procedure, and we find that it is advantageous to exclude the
large $k$ values for fitting the simulated $P^s_{\HI}(\kpe, \kpa)$.
The values of $(\kpe,\kpa)$ were restricted to be within
$\{0.9,1.1,1.5,1.7,2.0\} \, {\rm Mpc}^{-1}$ for the redshifts
$\{1,1.5,2,2.5,\geq 3\}$ respectively.

%----------------------------------------------------
\begin{table*}
	\centering
	{\renewcommand{\arraystretch}{1.5}%
		\begin{tabular}{|c|c|c|c|c|c|c|c|c|c|}
			\hline
			\multirow{2}{*}{$z$} & \multicolumn{3}{|c|}{$\tilde{b}(k)$} & \multicolumn{3}{|c|}{$b(k)$} & \multicolumn{3}{|c|}{$b_0$} \\
			\cline{2-10}
			& A1 & A2 & A3 & B1 & B2 & B3 & C1 & C2 & C3 \\
			\hline
			\multirow{2}{*}{1.0} & 0.904 & 1.479 & 1.102 & 0.846 & 1.734 & 1.135 & 3.157 & 6.792 & 4.515 \\
			& (1.080) & (1.557) & (1.025) & (1.008) & (1.784) & (0.959) & (1.535) & (5.692) & (2.615)\\
			\hline
			
			\multirow{2}{*}{1.5} & 0.935 & 1.486 & 1.139 & 0.915 & 1.754 & 1.214 & 3.408 & 6.756 & 4.747 \\
			& (0.912) & (2.241) & (0.946) & (0.899) & (2.629) & (0.927) & (1.459) & (7.377) & (2.797)\\
			\hline
			
			\multirow{2}{*}{2.0} & 0.709 & 1.190 & 0.883 & 0.731 & 1.428 & 0.978 & 1.884 & 3.898 & 2.674 \\
			& (1.112) & (2.910) & (0.744) & (1.183) & (3.375) & (0.759) & (0.938) & (6.641) & (1.590) \\
			\hline
			
			\multirow{2}{*}{2.5} & 0.639 & 1.114 & 0.829 & 0.672 & 1.308 & 0.920 & 0.792 & 1.604 & 1.117 \\
			& (0.689) & (5.840) & (1.109) & (0.743) & (6.469) & (1.143) & (0.618) & (7.356) & (1.358)\\
			\hline
			
			\multirow{2}{*}{3.0} & 0.835 & 1.178 & 0.979 & 0.908 & 1.377 & 1.104 & 1.637 & 1.676 & 1.646 \\
			& (0.678) & (7.024) & (1.216) & (0.739) & (7.575) & (1.243) & (1.649) & (4.927) & (1.252) \\
			\hline

			\multirow{2}{*}{3.5} & 0.371 & 0.439 & 0.399 & 0.379 & 0.481 & 0.421 & 3.270 & 3.259 & 3.265 \\
			& (0.625) & (4.678) & (1.119) & (0.637) & (4.977) & (1.131) & (2.919) & (2.660) & (2.137) \\
			\hline

			\multirow{2}{*}{4.0} & 0.465 & 0.475 & 0.469 & 0.454 & 0.472 & 0.462 & 8.004 & 8.004 & 8.004 \\
			& (0.665) & (3.075) & (1.122) & (0.642) & (3.241) & (1.120) & (4.805) & (3.691) & (4.124) \\
			\hline
			
			\multirow{2}{*}{4.5} & 0.649 & 0.647 & 0.648 & 0.608 & 0.604 & 0.606 & 15.339 & 15.339 & 15.339 \\
			& (0.774) & (1.607) & (0.918) & (0.744) & (1.669) & (0.905) & (6.167) & (5.732) & (5.945) \\
			\hline

			\multirow{2}{*}{5.0} & 1.022 & 1.022 & 1.022 & 0.956 & 0.956 & 0.956 & 26.067 & 26.067 & 26.067 \\
			& (1.271) & (1.836) & (1.424) & (1.216) & (1.842) & (1.384) & (10.974) & (10.913) & (10.944) \\
			\hline

			\multirow{2}{*}{5.5} & 1.786 & 1.786 & 1.786 & 1.542 & 1.542 & 1.542 & 34.162 & 34.162  & 34.162 \\
			& (1.279) & (1.381) & (1.287) & (1.219) & (1.350) & (1.235) & (14.909) & (14.909) & (14.909) \\
			\hline
			
			\multirow{2}{*}{6.0} & 2.312 & 2.312 & 2.312 & 2.018 & 2.018 & 2.018 & 43.866 & 43.866 & 43.866 \\
			& (1.152) & (1.145) & (1.139) & (1.089) & (1.096) & (1.081) & (23.944) & (23.944) & (23.944) \\	
			
			\hline
		\end{tabular}}
		\caption{This tabulates the minimum value of the reduced $\chi^2$ $(\chi^2/N)$ for the different models, here $N$ is the degrees of freedom. The $(\chi^2/N)$ values outside (inside the brackets) are related to the HC (HP) method.} 
		\label{tab:chisq}
		
	\end{table*}	
%----------------------------------------------------

%----------------------------------------------------

\begin{table}
	\centering
	{\renewcommand{\arraystretch}{1.5}%
		\begin{tabular}{|c|c|c|c|c|}
			\hline
			\multirow{3}{*}{$z$} & \multicolumn{2}{|c|}{HC} & \multicolumn{2}{|c|}{HP} \\
			\cline{2-5}
			&    A1       &     B1     &     A1      &     B1     \\
			\cline{2-5}
			& $\sigma_p$ (Mpc) & $\sigma_p$ (Mpc) & $\sigma_p$ (Mpc) & $\sigma_p$ (Mpc) \\
			\hline
			1.0	& $2.967 ^{+0.029} _{-0.029}$ & $3.359 ^{+0.030} _{-0.029}$ & $3.999 ^{+0.027} _{-0.026}$ & $4.431 ^{+0.028} _{-0.028}$          \\ \hline
			1.5	& $1.934 ^{+0.013} _{-0.013}$ & $2.142 ^{+0.014} _{-0.013}$ & $3.199 ^{+0.014} _{-0.014}$ & $3.448 ^{+0.015} _{-0.014}$          \\ \hline
			2.0	& $1.359 ^{+0.008} _{-0.008}$ & $1.475 ^{+0.008} _{-0.008}$ & $2.626 ^{+0.009} _{-0.008}$ & $2.764 ^{+0.008} _{-0.008}$          \\ \hline
			2.5	& $1.025 ^{+0.006} _{-0.005}$ & $1.096 ^{+0.006} _{-0.006}$ & $2.144 ^{+0.005} _{-0.006}$ & $2.222 ^{+0.006} _{-0.005}$          \\ \hline
			3.0	& $0.689 ^{+0.003} _{-0.004}$ & $0.733 ^{+0.004} _{-0.003}$ & $1.823 ^{+0.004} _{-0.004}$ & $1.876 ^{+0.004} _{-0.005}$          \\ \hline
			3.5	& $0.555 ^{+0.004} _{-0.005}$ & $0.591 ^{+0.004} _{-0.005}$ & $1.560 ^{+0.004} _{-0.004}$ & $1.595 ^{+0.004} _{-0.004}$          \\ \hline
			4.0	& $0.415 ^{+0.005} _{-0.005}$ & $0.448 ^{+0.004} _{-0.005}$ & $1.304 ^{+0.003} _{-0.004}$ & $1.331 ^{+0.003} _{-0.004}$          \\ \hline
			4.5	& $0.255 ^{+0.007} _{-0.007}$ & $0.292 ^{+0.007} _{-0.007}$ & $1.091 ^{+0.004} _{-0.004}$ & $1.113 ^{+0.004} _{-0.004}$          \\ \hline
			5.0	& $0.0 ^{+0.035} _{}$ &  $0.111 ^{+0.013} _{-0.015}$ & $0.934 ^{+0.004} _{-0.003}$ &  $0.953 ^{+0.004} _{-0.003}$          \\ \hline
			5.5	& $0.0 ^{+0.009} _{}$ & $0.0 ^{+0.010} _{}$ & $0.791 ^{+0.004} _{-0.003}$ & $0.809 ^{+0.003} _{-0.004}$          \\ \hline
			6.0	& $0.0 ^{+0.008} _{}$ &  $0.0 ^{+0.008} _{}$ & $0.650 ^{+0.004} _{-0.005}$ &  $0.667 ^{+0.005} _{-0.004}$ \\ \hline
		\end{tabular}}
		\caption{For models A1 and B1 (for the two \HI assignment methods), this presents the  best fit
			values of $\sigma_p$ along with the $1-\sigma$ uncertainty
			intervals at different redshifts.}
		\label{tab:best_fit_sigma}
	\end{table}
	
%----------------------------------------------------

%----------------------------------------------
\begin{figure}
	
	\psfrag{sigmap}[c][c][1.5][0]{$\sigma_p$ (Mpc)}
	\psfrag{z}[c][c][1.5][0]{$z$}
	\psfrag{A1}[c][c][1.5][0]{A1}
	\psfrag{B1}[c][c][1.5][0]{B1}
	\psfrag{HP}[c][c][1.5][0]{HP}
	\psfrag{HC}[c][c][1.5][0]{HC}	
	
	\centering
	\includegraphics[width=0.45\textwidth,angle=-90]{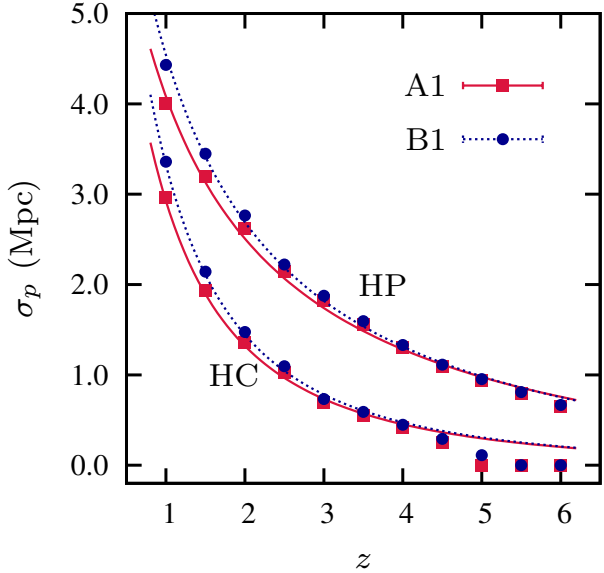}

	\caption{The red squares and the blue circles show the best
          fit values of $\sigma_p$ at different redshifts for models A1 and B1
          respectively (Table~\ref{tab:best_fit_sigma}). The solid red line and
          the dotted blue line show the predictions of eq.~(\ref{eq:sigma_z}). 
          The above two curves correspond to the HP method while the two 
          	curves below correspond to the HC method. }
	\label{fig:sigma_p_best_fit_models}
\end{figure}
%----------------------------------------

\section{Results}
\label{sec:results}

In this work we have modelled the anisotropy in the simulated
$P^s_{\HI}(\kpe,\kpa)$ using the nine models mentioned above
(eq.~(\ref{eq:modified_kaiser_model})).  Each model has a single free
parameter $\sigma_p$ to fit.  We perform a $\chi^2$ minimization with
respect to $\sigma_p$ to determine the value that best fits the
simulated $P^s_{\HI}(\kpe,\kpa)$. Table~\ref{tab:chisq} shows the
goodness of fit which is quantified by the reduced chi square
$\chi^2/N$ where $N$ is the degree of freedom. We consider models with
$\chi^2/N \sim 1$ are acceptable fits to the simulated
$P^s_{\HI}(\kpe,\kpa)$, and models with $\chi^2/N$ considerably larger
than unity are rejected.

We first discuss our results for the HC method. Considering the $\chi^2/N$
 values in Table~\ref{tab:chisq}, we find
that in the redshift range $z=2-3$ and possibly at $3.5$, nearly all
the models fit the simulated $P^s_{\HI}(\kpe,\kpa)$ reasonably well.
At higher redshifts ($z \ge 3.5$) we see that the constant bias models
(C1, C2 and C3) fail to fit the simulations.  We also see that at
lower redshifts ($z<2$), the constant bias models, particularly with
the Gaussian (C2) and the Lorentzian squared (C3) damping profiles, do
not provide a good fit.  On the other hand, the scale dependent
complex bias models (A1, A2 and A3) and the scale dependent real bias
models (B1, B2 and B3) provide reasonably good fits to the simulated
$P^s_{\HI}(\kpe,\kpa)$ through nearly the entire $z$ range considered
here.  However, at $z \geq 5.5$, the $\chi^2/N$ values are greater
than $1$ for the scale dependent complex and real bias models.
Now we discuss the results for the HP method. Almost all the above 
conclusions hold true for the HP method. However, for $z=2-4$, models A2 
and B2 show large $\chi^2/N$ ($>1$) and hence do not give a good fit to the simulated
$P^s_{\HI}(\kpe,\kpa)$. At $z \ge 5.5$, the $\chi^2/N$ $\sim 1$ (less than
 the values obtained for the HC method) indicative of a better fit to the simulations.

Considering the $\chi^2/N$ values in the entire $z$ range, we find
that the Lorentzian damping profile provides the best fit compared to
the two other profiles.  The Lorentzian squared profile performs
better than the Gaussian profile, but it has slightly higher
$\chi^2/N$ values compared to the Lorentzian profile 
at some of the reddshifts.  All the three damping profiles,
 combined with the scale dependent real and complex bias models,
 work almost equally well in the range $z=2-5$ for the HC method. 
In case of the HP method, the Lorentzian and the Lorentzian squared
 profiles combined with the scale dependent real and complex 
bias models are comparably good in the entire redshift range. 
The Gaussian damping profile show somewhat larger $\chi^2/N$ at
 the intermediate redshifts for the HP method. 
 
We can broadly summarize that the Lorentzian damping profile coupled
with the scale dependent complex bias and the scale dependent real
bias (models A1 and B1) both work nearly equally well and they best
fit the simulated power spectrum over the entire $z$ range considered
here. We mostly focus on these two models in the subsequent discussion
of this paper.

Table~\ref{tab:best_fit_sigma} provides the best fit values of
$\sigma_p$ at different $z$ for the two models A1 and B1 
for the two different \HI mass assignment methods.  Here
$\sigma_p$ is in units of comoving Mpc, or equivalently $[\sigma_p a
H(a)]$ in units of km/s. For the HC method, we see that the best fit values of
$\sigma_p$ decrease with increasing redshift and we have $\sigma_p
\sim 0$ for $z \geq 5$ for Model A1 and for $z > 5$ for Model B1 which
implies that the FoG effect is not very prominent at the higher
redshifts.  This also indicates that the FoG damping can safely be
ignored for $z>5$, and the anisotropy in $P^s_{\HI}(\kpe,\kpa)$ can be
adequately modelled by only using the modified Kaiser term with a
scale dependent complex or real bias if one uses the HC method. 
For the HP method, the best fit values of $\sigma_p$ also decrease with
 increasing redshift. However, unlike the HC method, here the best 
fit values of $\sigma_p$ are non zero even at the highest redshift. 
The $\sigma_p$ values are larger than the HC method and this indicates 
that the HP method takes into account the FoG effect due to the
 velocity dispersion inside the haloes which is ignored by the HC method.

Figure~\ref{fig:sigma_p_best_fit_models} shows the best fit values of
$\sigma_p$ as a function of redshift for the two models A1 and B1. For
both the models $\sigma_p$ shows a nearly parabolic $z$ dependence for
$z<5$ ($z \le 6$) for the HC (HP) method.  We have used
the following functional form to fit the $z$ dependence of $\sigma_p $,
%---------------------------------------
\begin{equation}
\sigma_p(z)=\sigma_{p}(0) \,  (1+z)^{-m} \exp \left[ - \left(
  \frac{z}{z_p} \right) ^2 \right]\,, 
\label{eq:sigma_z}
\end{equation} 
%---------------------------------------
where $\sigma_{p}(0)$, $m$ and $z_p$ are three fitting parameters.  
Considering the HC method, we have carried out a rough fitting 
for Model A1 (B1) with the values of the parameters 
$\sigma_{p}(0)=$11.0 (13.4) Mpc, $m=$ 1.9 (2.0) and
$z_p=$11.0 (11.5) which works in the range
$1 \le z <5$. The same fitting for the HP method yields
$\sigma_{p}(0)=$9.12 (10.8) Mpc, $m=$ 1.15 (1.24) and $z_p=$12.0
(12.0) for Model A1 (B1) which works in the range $1 \le z < 6$.  We
see that the values of $\sigma_p$ are slightly larger for Model B1 as
compared to Model A1, however it is not clear if the differences are
statistically significant.  The $\sigma_p$ values are also higher
for the HP method which again indicates that the FoG damping
increases when the velocity dispersion inside the haloes is taken into
account.  However, $\sigma_p$ falls slowly with redshift for the HP
method in comparison to the HC method. Considering the HC method, our
fit fails for $z \geq 5$ where the $\sigma_p$ values lie below the fit
and are consistent with zero.  On the other hand, the fit works
almost for the entire $z$ range for the HP method.

%--------------------------------------------------
\begin{figure*}
	
  \psfrag{z=1}[c][c][0.8][0]{$z=1\,$}
  \psfrag{z=2}[c][c][0.8][0]{$\quad\;\;2$}
  \psfrag{z=3}[c][c][0.8][0]{$\quad\;\;3$}
  \psfrag{z=6}[c][c][0.8][0]{$\quad\;\;6$}
  
  \psfrag{zz=1}[c][c][0.75][0]{$z=1\,$}
  \psfrag{zz=2}[c][c][0.8][0]{$\;2$}
  \psfrag{zz=3}[c][c][0.8][0]{$\;3$}
  \psfrag{zz=6}[c][c][0.8][0]{$6\quad$}
  
  \psfrag{z2=1}[c][c][0.8][0]{$z=1\quad$}
  \psfrag{z2=2}[c][c][0.8][0]{$\;\;2$}
  \psfrag{z2=3}[c][c][0.8][0]{$\;\;3$}
  \psfrag{z2=6}[c][c][0.8][0]{$\;\;6$}
  
  \psfrag{HI Real}[c][c][1.3][0]{$\Delta^{2}_{\HI}(k)$}
  \psfrag{HI monopole}[c][c][1.3][0]{$\Delta^{2s}_{0}(k)$}
  \psfrag{HI quadrupole}[c][c][1.3][0]{$\Delta^{2s}_{2}(k)$}
  
  \psfrag{HC}[c][c][0.9][0]{HC}
  \psfrag{HP}[c][c][0.9][0]{HP}

  \psfrag{k}[c][c][1.4][0]{$k~{\rm Mpc}^{-1}$}

  \centering
  \includegraphics[width=0.33\textwidth,angle=-90]{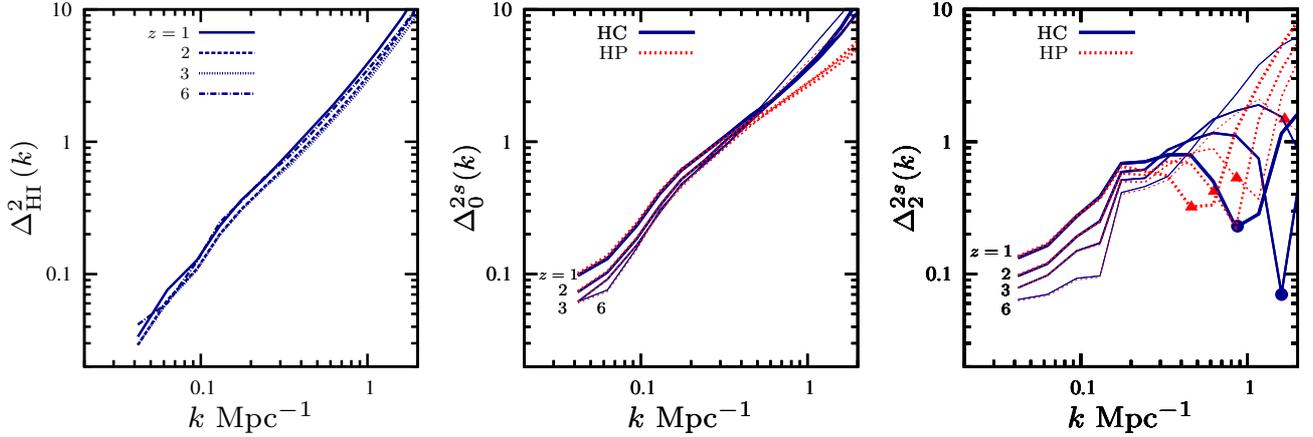}
  \caption{Each panel here shows the dimensionless power spectrum
    $\Delta^2(k)=k^3 P(k)/ 2 \pi^2$ at the four indicated $z$ values.
    The left panel shows results for $P_{\HI}(k)$. The center and right
    panels respectively show results for $P^s_0(k)$ and $P^s_2(k)$
    considering both the HC and HP methods as indicated in the
    figure. Note that the quadrupole $P^s_2(k)$ becomes negative at large
    $k$, this transition is marked by a circle (triangle) for the HC (HP)
    method. For the HC method this transition occurs at $k > 2 {\rm
      Mpc}^{-1}$ for $z \ge 3$. }
  \label{fig:combined_multipoles}
\end{figure*}
%-----------------------------------------------------

It is convenient to decompose the anisotropic redshift space \HI power
spectrum $P^s_{\HI}(\kpe,\kpa)$ into angular multipoles
\citep{hamilton-92-RSDcorrelation,cole-fisher-weinberg94} 
%----------------------------------------------------
\begin{equation}
P^s_{\HI}(k, \mu)= \sum\limits_{\ell}^{} \mathcal{L}_{\ell}(\mu) P^s_{\ell}(k) \,,
\label{eq:legend_moments}
\end{equation}
%------------------------------------------------------
where $\mathcal{L}_{\ell}(\mu)$ are the Legendre polynomials and
$P^s_{\ell}(k)$ are the different angular moments of the power
spectrum.  The angular moments $P^s_{\ell}(k)$ are functions of a
single variable $k$ and therefore are relatively easy to visualize and
interpret.  Only the even moments survive in the flat sky
approximation.  Further, in linear theory ({\textit i.e.} with just
the Kaiser enhancement) only the first three even moments $\ell = 0$
(monopole), $2$ (quadrupole) and $4$ (hexadecapole) are non-zero.
Here we have calculated the first three even moments from the
simulated $P^s_{\HI}(\kpe,\kpa)$. The hexadecapole ($P^s_{4}(k)$) is
found to be rather noisy (large cosmic variance), and we have included
results only for the monopole $P^s_{0}(k)$ and the quadrupole
$P^s_{2}(k)$. We have compared these to the predictions of the
different models in order to analyze how well the models are able to
reproduce the simulated \HI power spectrum.

 Considering both the HC and HP methods,
  Figure~\ref{fig:combined_multipoles} shows the dimensionless power
  spectra $\Delta^{2s}_0 (k) =k^3 P^s_0(k) /2 \pi^2$ and $\Delta^{2s}_2 (k)=k^3
  P^s_2(k)/2 \pi^2$. The real space \HI power spectrum
  $\Delta^2_{\HI}(k)=k^3 P_{\HI}(k)/ 2 \pi^2$ is also shown for
  reference. Unlike the real space dark matter power spectrum  $P(k)$
  which evolves as  $(1+z)^{-2}$ in the linear regime,  
  we see that the real space \HI power spectrum shows a  very 
  weak redshift evolution (see Paper I for a detailed discussion).
  However,  we see that both  $\Delta^{2s}_{0}(k)$ and
 $\Delta^{2s}_{2}(k)$ evolve  with redshift, the effect being
  relatively more pronounced for $\Delta^{2s}_{2}(k)$.
  We first consider the large length-scales    (small $k$) where 
  the results   from the HC and HP  methods are indistinguishable.
  We see that both  $\Delta^{2s}_0$ and $\Delta^{2s}_2$ grow with
  decreasing $z$. This growth is due to   the RSD which  occurs here 
  primarily due to the coherent flows. This component  of RSD is
  expected to be  identical in both the methods which is why their
  results are indistinguishable.  The monopole,
  which  has contributions from both $P_{\HI}(k)$ and the RSD,
  does not show very significant evolution at large $z$ $(z> 3)$
  and shows a modest growth only at low $z$  ($z \le3$). This can
  be explained by noting that the bias $b(z)$ increases with $z$
  and has value $b(z)>2$ for $z>3$ whereby $\beta$, which determines
  the relative contribution from the RSD
  (eq.~\ref{eq:modified_kaiser_model}), has a very small value at
  lagre $z$. The quadrupole, which arises  entirely due to the RSD,
  appears to evolve through the entire $z$ range.

Considering the small length-scales (large $k$), we find that the
results from the two methods are different.  The RSD here is primarily
due to the FoG suppression arising form the velocity dispersion which
is different for the two methods.  The HP method has an enhanced FoG
suppression and the monopole $\Delta^{2s}_{0}(k)$ is smaller than that
for the HC method.  Further, for both the methods the $z$ evolution is
opposite to that seen at small $k$, and the value of
$\Delta^{2s}_{0}(k)$ is found to decrease with decreasing $z$.  We can
explain this by noting that the velocity dispersion and the FoG
suppression increases with decreasing redshift
(Figure~\ref{fig:sigma_p_best_fit_models}). For both the methods, the
quadrupole is negative at large $k$. This essentially arises due to
the FoG elongation of structures along the line of sight direction.
We find that the $k$ value corresponding to this transition decreases
with decreasing $z$. Further, these $k$ values are relatively larger
for the HC method where the FoG is restricted to relatively smaller
length-scales.

%--------------------------------------------------
\begin{figure*}
	
  \psfrag{z=1.0}[c][c][1.5][0]{$z=1\quad$}
  \psfrag{z=2.0}[c][c][1.5][0]{$z=2\quad$}
  \psfrag{z=3.0}[c][c][1.5][0]{$z=3\quad$}
  \psfrag{z=4.0}[c][c][1.5][0]{$z=4\quad$}
  \psfrag{z=5.0}[c][c][1.5][0]{$z=5\quad$}
  \psfrag{z=6.0}[c][c][1.5][0]{$z=6\quad$}
  \psfrag{HI monopole ratio}[c][c][1.5][0]{$P^s_0(k)/ P^s_{0\,, B1}(k)$}
  
  \psfrag{Simulations}[c][c][0.9][0]{Simulations$\quad$}
  \psfrag{A1}[c][c][0.9][0]{A1}
  \psfrag{B1}[c][c][0.9][0]{B1}
  \psfrag{B2}[c][c][0.9][0]{B2}
  \psfrag{B3}[c][c][0.9][0]{B3}
  \psfrag{C1}[c][c][0.9][0]{C1}
  \psfrag{A2}[c][c][0.9][0]{A2}
  
  \psfrag{kmode}[c][c][1.6][0]{$k~{\rm Mpc}^{-1}$}
  
  \psfrag{HC}[c][c][1.3][0]{HC}
  \psfrag{HP}[c][c][1.3][0]{HP}

  \centering
  \includegraphics[width=0.52\textwidth,angle=-90]{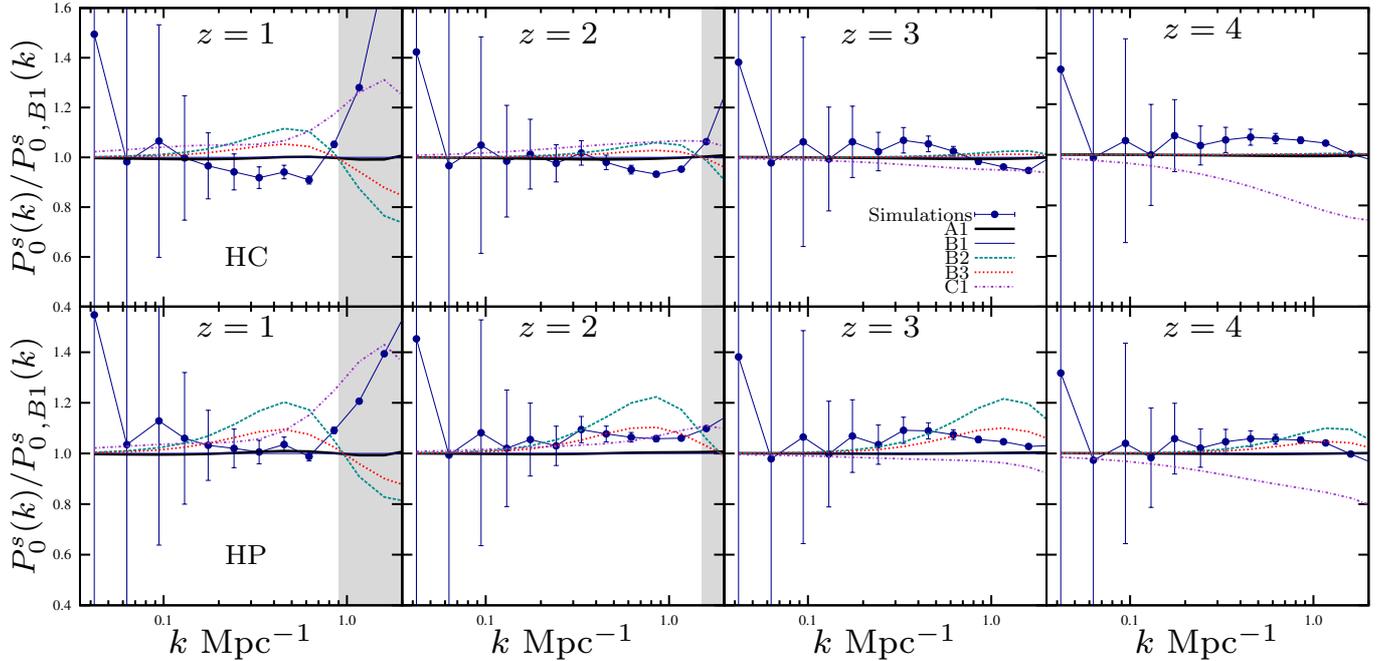}

  \caption{This shows the simulated and model monopole $P^s_0 (k)$
    of the redshift space \HI power spectrum for 
   the HC (HP) method in the upper (lower) panels. The values shown
    here are  normalized by the results for     Model B1. The blue
    circles (connected with blue solid line) and the 
    vertical error bars present the mean and the $1-\sigma$ spread
    determined from the five independent realizations of the
    simulations. The other lines correspond to different model
    predictions as indicated in the figure legend. The shaded regions
    denote the $k$ 
    range which  has  not been used for the fitting.}
  \label{fig:HI_monopole}
\end{figure*}
%-----------------------------------------------------

Considering the monopole $P^s_0(k)$, the upper panels of
Figure~\ref{fig:HI_monopole} show a comparison of the model
predictions against the values obtained from the simulations for the
HC method. For convenience, the values of $P^s_0(k)$ have been
normalized using the predictions for Model B1. The lower panels of
Figure~\ref{fig:HI_monopole} show the same quantities for the HP
method.  We see that the results from the two methods are
qualitatively similar over a large portion of the $k$ range across all
the redshifts shown here, and as noted earlier the results are nearly
indistinguishable at small $k$ ($< 0.2 \, {\rm Mpc}^{-1}$). At $z=1$,
models A1 and B1 match the simulated values for $k < 0.3 \, {\rm
Mpc}^{-1}$ for the HC method while this range increases to $k < 0.6 \,
{\rm Mpc}^{-1}$ for the HP method. For both the methods, the
deviations of A1 and B1 from the simulated values lie within $10 \%$
through the entire fitting range.  All the other models deviate from
the simulations at a smaller $k$ value ($k \sim 0.2 \, {\rm
Mpc}^{-1}$). The deviations from the simulations are also larger
compared to models A1 and B1. The behaviour at $z=2$ for the HC method
is similar to that seen at $z=1$.  Models A1 and B1 match the
simulations up to $k < 0.5 \, {\rm Mpc}^{-1}$, whereas all the other
models match the simulations for a smaller $k$ range ($k < 0.4 \, {\rm
Mpc}^{-1}$).  For models A1 and B1, the deviations from the
simulations are within $10\%$ across the entire fitting range,
whereas we find up to $\sim 15 \%$ deviations for all the other
models. In contrast, for the HP method Model B3 provides a better fit
compared to the models A1, B1 and all the other models. Models A1 and
B1 match the simulated values within $k \sim 0.2 \,{\rm Mpc}^{-1}$
while this extends to $k < 0.4 \,{\rm Mpc}^{-1}$ for Model B3.  The
deviations from the simulations are within $10\%$ across the entire
fitting range for all three of these models.  At $z=3$, in case of the
HC method almost all the models agree well with the simulations for $k
< 0.3 \, {\rm Mpc}^{-1}$ . Further, the deviations from the
simulations are within $15\%$ through the entire $k$ range. For the HP
method, Model B2 matches the simulations for a larger $k$ range ($k <
0.5 \, {\rm Mpc}^{-1}$) compared to the other models which deviate
after $k \sim 0.2 \, {\rm Mpc}^{-1}$. However, the deviations from the
simulations are within $10\%$ for all the models barring B2 and C1. At
$z= 4$ the models with a scale independent bias all fail to match the
simulated monopole, and for the monopole we focus only on the models
with a $k$ dependent bias, either complex or real.  Considering the HC
method, the predictions of all these models are indistinguishable
across the entire $k$ range irrespective of the damping profile, and
they match the simulations for $k < 0.3 \,{\rm Mpc}^{-1}$. Considering
the HP method which has a larger velocity dispersion, the predictions
depend on the damping profile and Model B2 fits the simulations over a
relatively large $k$ range $k < 0.6 \,{\rm Mpc}^{-1}$ whereas the
other damping profiles work well within $k < 0.3 \,{\rm Mpc}^{-1}$.
For both the methods, the deviations from the simulations are within
$\sim 10 \%$ over the entire fitting range. The predictions at $z \geq
5$ are very similar to those at $z=4$ and these have not been shown
separately in Figure~\ref{fig:HI_monopole} and any of the subsequent
figures.  We find that the models with a scale dependent bias are able
to match the simulations at $k < 0.3 \,{\rm Mpc}^{-1}$ and $k < 0.4
\,{\rm Mpc}^{-1}$ for the HC and HP methods respectively.  For the HC
method the predictions for the different damping profiles are
indistinguishable, whereas there are small differences (which get
smaller with increasing $z$) for the HP method.

%-----------------------------------------------------
\begin{figure*}
	
	\psfrag{z=1.0}[c][c][1.5][0]{$z=1\:$}
	\psfrag{z=2.0}[c][c][1.5][0]{$z=2\:$}
	\psfrag{z=3.0}[c][c][1.5][0]{$z=3\:$}
	\psfrag{z=4.0}[c][c][1.5][0]{$z=4\:$}
	\psfrag{z=5.0}[c][c][1.5][0]{$z=5\:$}
	\psfrag{z=6.0}[c][c][1.5][0]{$z=6\:$}
	\psfrag{HI quadrupole ratio}[c][c][1.5][0]{$P^s_2(k)/ P^s_{2\,, B1}(k)$}

	\psfrag{Simulations}[c][c][0.9][0]{Simulations$\quad$}
	\psfrag{A1}[c][c][0.9][0]{A1}
	\psfrag{B1}[c][c][0.9][0]{B1}
	\psfrag{B2}[c][c][0.9][0]{B2}
	\psfrag{B3}[c][c][0.9][0]{B3}
	\psfrag{C1}[c][c][0.9][0]{C1}
	\psfrag{A2}[c][c][0.9][0]{A2}
	
	\psfrag{Relative error}[c][c][1.2][0]{$\delta P^s_0/P^s_0$}
	\psfrag{kmode}[c][c][1.6][0]{$k~{\rm Mpc}^{-1}$}
	
   \psfrag{HC}[c][c][1.3][0]{HC}
   \psfrag{HP}[c][c][1.3][0]{HP}

	\centering
	\includegraphics[width=0.52\textwidth,angle=-90]{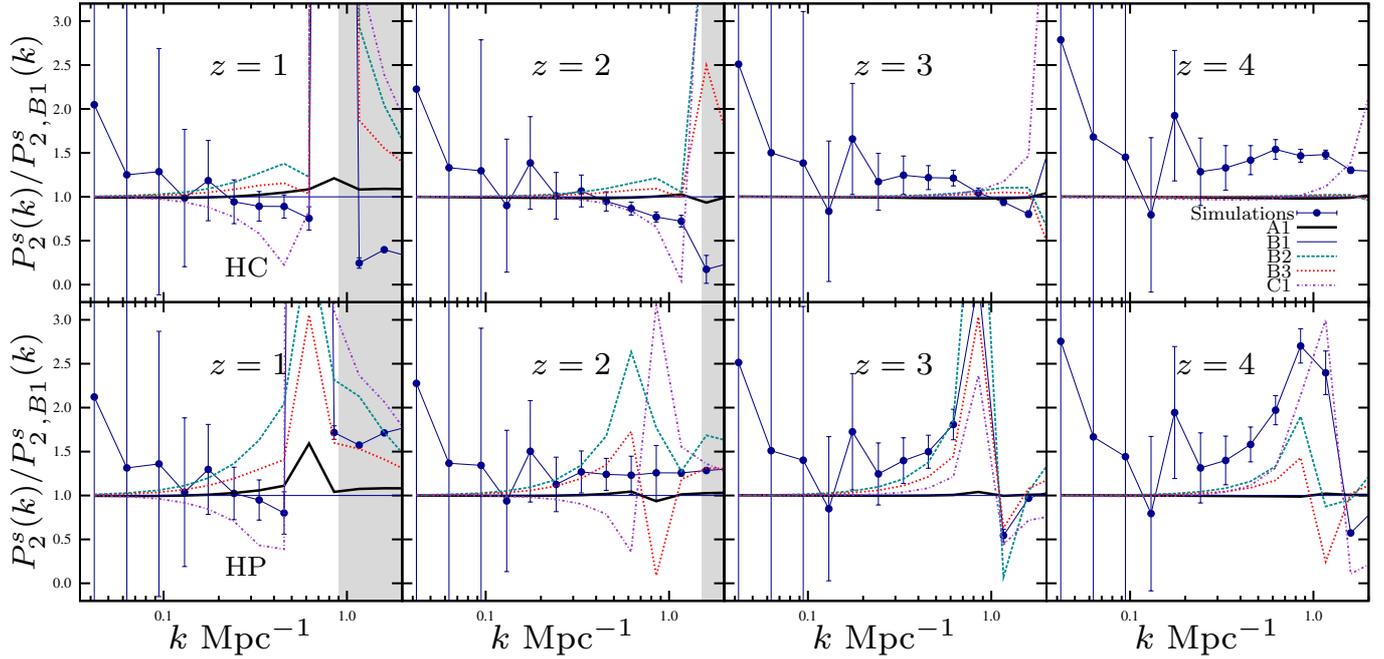}

	\caption{Same as Figure~\ref{fig:HI_monopole} for the
          quadrupole $P^s_2 (k)$.} 
	\label{fig:HI_quadrupole}
\end{figure*}
%-----------------------------------------------------

%-----------------------------------------------------

Figure~\ref{fig:HI_quadrupole} shows $P^s_2 (k)$, the results are
normalized by Model B1.  We see that at $z=1$, models A1 and B1
perform better than the other models in matching the simulations for
both the HC and HP methods.  The match extends to $k < 0.5 \, {\rm
Mpc}^{-1}$ and $k < 0.3 \, {\rm Mpc}^{-1}$ for the HC and HP methods
respectively, and the deviations are within $20 \%$ at $k \lesssim 0.5
\, {\rm Mpc}^{-1}$ for the HP method.  For both the methods, these
models do not match the simulated quadrupole beyond this $k$ range,
and some of the simulated values are in excess of the range considered
in the figure.  At $z=2$ we see that all the models with $k$ dependent
bias match the simulations up to $k < 0.5 \,{\rm Mpc}^{-1}$ for the HC
method while only Model B3 is able to cover the same range for the HP
method where the other scale dependent bias models match the
simulations only to $k < 0.3 \,{\rm Mpc}^{-1}$.  Surprisingly Model C1
matches the simulations for a large $k$ range ($k < 0.7 \, {\rm
Mpc}^{-1}$) for the HC method, however this is reduced to ($k < 0.3 \,
{\rm Mpc}^{-1}$) for the HP method.  For both the methods, across the
entire fitting range the deviations from the simulated values are
within $30\%$ for models A1 and B1, while all the other models show
somewhat larger deviations.  At $z=3$ for the HC method we see that all
the models match the simulations for $k < 0.3 \, {\rm Mpc}^{-1}$, and
for models A1 and B1 the deviations lie within $20\%$ across the
entire $k$ range while these are somewhat larger for the other
models. For the HP method Model B2 matches the simulations for a
relatively larger $k$ range ($ \lesssim 0.6 \,{\rm Mpc}^{-1}$) while
this is restricted to $k < 0.3 \,{\rm Mpc}^{-1}$ for the other models.
All the models deviate significantly from the simulations at large
$k$.  At $z=4$ for both the methods the models match the simulated
values for a very limited range $k < 0.15 \, {\rm Mpc}^{-1}$ and
differ significantly beyond this. At $z>4$ the results are very
similar to those for $z=4$ and we have not shown these here.

%-----------------------------------------------------
\begin{figure*}
	
	\psfrag{z=1.0}[c][c][1.5][0]{$\, z=1$}
	\psfrag{z=2.0}[c][c][1.5][0]{$\, z=2$}
	\psfrag{z=3.0}[c][c][1.5][0]{$\, z=3$}
	\psfrag{z=4.0}[c][c][1.5][0]{$\, z=4$}
	\psfrag{z=5.0}[c][c][1.5][0]{$\, z=5$}
	\psfrag{z=6.0}[c][c][1.5][0]{$\, z=6$}

	\psfrag{Simulations}[c][c][0.8][0]{Simulations$\,\,$}
	\psfrag{A1}[c][c][0.9][0]{A1}
	\psfrag{B1}[c][c][0.9][0]{B1}
	\psfrag{B2}[c][c][0.9][0]{B2}
	\psfrag{B3}[c][c][0.9][0]{B3}
	\psfrag{C1}[c][c][0.9][0]{C1}
	\psfrag{A2}[c][c][0.9][0]{A2}
	\psfrag{Linear Theory}[c][c][0.8][0]{Linear Theory$\,\,$}
	
	\psfrag{quadpole by monopole}[c][c][1.2][0]{$P^s_2(k)/P^s_0(k)$}
	\psfrag{kmode}[c][c][1.2][0]{$k~{\rm Mpc}^{-1}$}
	
   \psfrag{HC}[c][c][1.3][0]{HC}
   \psfrag{HP}[c][c][1.3][0]{HP}

	\centering
	\includegraphics[width=0.52\textwidth,angle=-90]{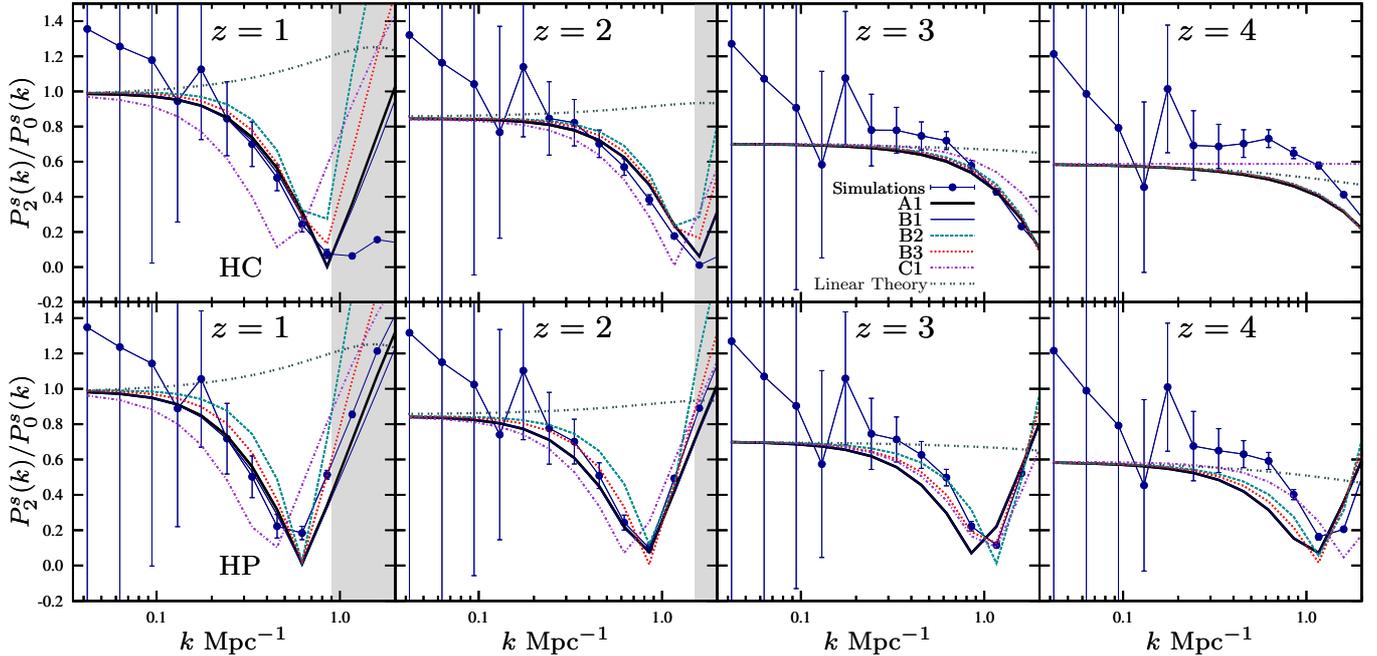}

	\caption{The ratio $P^s_2 (k)/P^s_0 (k)$ for the HC and HP methods are
          shown in the upper and lower panels respectively.  The blue circles
          (connected with blue solid line) and the vertical error bars,
          respectively, present the mean and the $1-\sigma$ spread determined
          from the five independent realizations of the simulations. Predictions
          from different models and linear theory are shown for comparison.  The
          shaded regions denote the $k$ range which has not been used for the
          fitting. Note that the ratio has negative values beyond the sharp dip
          at large $k$. } 
	\label{fig:HI_quadpole_by_monopole}
\end{figure*}
%-----------------------------------------------------

%-----------------------------------------------------

Considering the ratio $P^s_2 (k)/ P^s_0 (k)$, the upper (lower) panels of
Figure~\ref{fig:HI_quadpole_by_monopole} show the simulated values
along with the model predictions and also the linear theory prediction
for the HC (HP) method.  The linear theory prediction is computed
using just the Kaiser enhancement term with the scale dependent real
bias $b(k)$.  We see that, at $z=1$ models A1 and B1 match the
simulations for the entire fitting range for the HC method while this
range shrinks to $k < 0.6\, {\rm Mpc}^{-1}$ for the HP
method. Considering the HC method, models B2 and B3 match the
simulations for $k < 0.4 \, {\rm Mpc}^{-1}$ and $k < 0.7 \,{\rm
Mpc}^{-1}$ respectively while, for the HP method, they match the
simulations up to $k \sim 0.2 \, {\rm Mpc}^{-1}$ and $k \sim 0.3 \,
{\rm Mpc}^{-1}$ respectively. Model C1 and the linear theory
prediction deviate from the simulations before $k \sim 0.3 \, {\rm
Mpc}^{-1}$ ($k \sim 0.2 \, {\rm Mpc}^{-1}$) for the HC (HP) method.
At $z=2$, models A1 and B1 match the simulations for $k < 0.7 \, {\rm
Mpc}^{-1}$ for the HC method and this range increases to $k \sim 1 \,
{\rm Mpc}^{-1}$ for the HP method. All the other models deviate from
the simulations at a smaller $k$ value. The linear theory predictions
agree with the simulations for $k < 0.4 \, {\rm Mpc}^{-1}$ and $k <
0.3\, {\rm Mpc}^{-1}$ for the HC and HP methods respectively.  At
$z=3$, for the HC method, all the models with a scale dependent bias
are in reasonable agreement with the simulations over the entire $k$
range while models with a scale independent bias deviate at large $k$.
For the HP method Model B2, which works best, matches the simulations
for $k < 0.7 \, {\rm Mpc}^{-1}$ while this range is limited to $k
\sim 0.3 \, {\rm Mpc}^{-1}$ for all the other models. For both the
methods, the linear theory agrees with the simulations for $k < 0.5 \,
{\rm Mpc}^{-1}$.  At $z=4$, considering both the HC and HP methods,
the models with a scale dependent bias match the simulations for $k <
0.3 \, {\rm Mpc}^{-1}$. However, for both the methods, Model C1 which
has a scale independent bias and the linear theory prediction agree
with the simulations for $k < 0.4 \, {\rm Mpc}^{-1}$.  At $z>4$, for
both the HC and HP methods, all the model predictions along with the
linear theory match the simulations for a very limited $k$ range ($k <
0.15 \, {\rm Mpc}^{-1}$). The results are nearly similar to those at
$z=4$ and these have not been shown here.

\section{Summary and Discussion}
\label{sec:summary}

The post-reionization \HI 21-cm signal, which is expected to be a
pristine probe of the large scale structures in the Universe, is an
excellent candidate for precision cosmology. This requires accurate
and reliable modelling of the expected signal. In an earlier paper
(Paper I) we have simulated the expected \HI 21-cm power spectrum
$P_{\HI}(k)$ in real space (as against redshift space) and used this
to model the $k$ dependence of the (possibly complex) bias
$\tilde{b}(k)$ over the redshift range $1 \le z \le 6$.  In this paper
we have extended the earlier simulations to include the redshift space
distortion (RSD) due to the peculiar motion of the \HI, and we have
used this to model the anisotropy of the redshift space \HI 21-cm
power spectrum $P^s_{\HI}(\kpe,\kpa)$. Such modelling is important on
two counts. This is first required to make accurate predictions for
various instruments which aim to carry out precision cosmology
observations using the \HI 21-cm signal. Precise modelling is also
required to correctly interpret the signal and extract the relevant
astrophysical and cosmological informations once the \HI 21-cm signal
is measured.

Here we have used two separate methods, subsequent to the \HI
assignment scheme in eq.~\ref{eq:bagla_scheme}, to distribute \HI in
the haloes. In the first method we place  the \HI content of a halo 
 at the halo center of mass and we refer to this as the HC method. The
 whole analysis of Paper I is based on the HC method. In the second
 method we  distribute  the \HI content of a halo equally among all the
member particles of the halo and this is referred to as the HP
method. The  real space clustering of \HI in these two methods differ
very little  and the difference can be ignored on large
scales. However, this  is not true in redshift space where the  HC method 
suppresses the FoG damping  as it does not take into
account the velocity dispersion within a halo. On the other hand, the
HP  method which incorporates this component of the 
velocity dispersion possibly overestimates the FoG effect. 
It should be noted that the two methods are expected to have the same
coherent flows and differ only in the velocity dispersion. 
The actual \HI distribution is possibly somewhere in between the two
methods which we have considered. We interpret the results from
the two methods as representing two limiting cases, and we expect the
predictions for the actual \HI distribution to lie somewhere in
between.  The differences between 
the two methods are apparent from Figure~\ref{fig:HI_power_spectrum}
and Figure~\ref{fig:HI_power_spectrum_HP} where the
$P^s_{\HI}(\kpe,\kpa)$ obtained from two methods is plotted against
its real space counter part at different redshifts.

We have modelled $P^s_{\HI}(\kpe,\kpa)$ using the simple assumption
(eq.~(\ref{eq:modified_kaiser_model})) that this can be obtained by
multiplying the model predictions of Paper I $(P_{\HI}(k)=b^2 P(k))$
with a Kaiser enhancement term and a Finger of God (FoG) damping term.
The various models considered here
(Section~\ref{sec:model-power-spec}) all have a single free parameter
$\sigma_p$ which is the pair velocity dispersion that appears in the
FoG damping term.  We have carried out a $\chi^2$ minimization with
respect to $\sigma_p$ in order to determine how well our models match
the simulated $P^s_{\HI}(\kpe, \kpa)$. Using the $\chi^2/N$ values
(Table~\ref{tab:chisq}) to estimate the goodness of fit for both
 the \HI assignment methods, we find that
models A1 and B1, both of which have a Lorentzian damping profile with
a scale dependent bias (complex and real respectively), provide the
best match to the simulations over the entire $z$ range. The same two
bias schemes combined with a Gaussian (A2, B2) and Lorentzian squared
(A3, B3) damping also work reasonably well, with the Lorentzian squared
having lower $\chi^2/N$ compared to the Gaussian. In contrast, the
models with a scale independent bias (C1, C2, and C3) fail to match the
simulations. However,  we note that there is a
degradation in the goodness of fit with increasing $z$ at $z \ge 5$
for the HC method (Table~\ref{tab:chisq}), and our models do not provide a very good fit
to the simulations at $z \ge 5$ even if we incorporate a scale
dependent complex or real bias.

Considering the HC method, for models A1 and B1, we find that the
best fit values of $\sigma_p$ (Table~\ref{tab:best_fit_sigma} and
Figure~\ref{fig:sigma_p_best_fit_models}) decrease approximately as
$(1+z)^{-2}$ with increasing $z$ (eq.~(\ref{eq:sigma_z})), and the
values of $\sigma_p$ are consistent with zero for $z >5$ . This
essentially tells us that it is not necessary to include the FoG
effect for modelling $P^s_{\HI}(\kpe,\kpa)$ at $z >5$ if one uses
the HC method.  This is also clearly seen in the simulations
(Figure~\ref{fig:HI_power_spectrum}) where there is no evidence for
the FoG effect at $z \ge 5$.  Considering the HP method, we see
that the the best fit values of $\sigma_p$
(Table~\ref{tab:best_fit_sigma} and
Figure~\ref{fig:sigma_p_best_fit_models}) fall relatively slowly
($(1+z)^{-1.2}$) as compared to the HC method. Further, unlike in the
HC method, $\sigma_p$ has a small but finite value even at the highest
redshift. We may interpret this as arising from the velocity
dispersion within a halo. This is expected to lead to an enhanced FoG
suppression for the HP method as noticeable in
Figure~\ref{fig:HI_power_spectrum_HP}.  In contrast to the HC method,
for the HP method the $k$ dependent complex and real bias models
provide a reasonably good fit to the simulations
(Table~\ref{tab:chisq}) over the entire $z$ range.

The angular moments $P^s_{\ell}(k)$ are relatively easy to visualize
and interpret.  Considering the monopole $P^s_0(k)$ and quadrupole
$P^s_2(k)$, we have investigated how well our models match the
simulations. Considering both the HC and HP methods,
for $P^s_0(k)$ we find that models A1 and B1 match the simulations at
large scales  $(k < 0.3 \, {\rm Mpc}^{-1})$ over the entire $z$ range,
and the  deviations are within $10-20 \%$ for larger $k$ values within
the  fitting range (Figure~\ref{fig:HI_monopole}).  
 The deviations are somewhat larger for the other models with a scale
 dependent bias , whereas the scale independent bias models show
 significantly larger deviations  particularly at high $z$. 
For $P^s_2(k)$, models A1 and B1 match the  
simulations (Figure~\ref{fig:HI_quadrupole})
over a relatively large $k$ range $(k < 0.5 \, {\rm Mpc}^{-1}$ 
  for the HC method and   $k < 0.3 \, {\rm Mpc}^{-1}$ for the HP
  method) at low $z$ $( \le 2)$.  
The $k$ range shrinks to $k < 0.3 \,{\rm Mpc}^{-1}$ for the HC
  method and  $k < 0.2 \,{\rm Mpc}^{-1}$ for the  HP method at $z =
3$, and is even smaller $(k < 0.15 \,{\rm Mpc}^{-1}$) for both the
  methods at $z \ge 4$. We find that all the models considered here  
significantly underpredict the quadrupole at $z \ge 4$, and these
deviations increase with increasing $z$.

The linear theory of redshift space distortion predicts
\citep{hamilton-98-rsd-review}
%--------------------------------------------------------------------
\begin{equation} 
\frac{P^s_2(k)}{P^s_0(k)}=\frac{(4/3) \beta + (4/7) \beta^2}{1+(2/3)
  \beta + (1/5) \beta^2} \,.
\label{eq:ratio}
\end{equation}
%--------------------------------------------------------------------
Measuring this ratio from observations of the \HI 21-cm power spectrum
holds the promise of determining $\beta=f(\Omega)/b$.  Assuming that
the bias $b$ is known, this can be used to determine $f(\Omega)$ the
growth rate of density perturbations which is a sensitive probe of
cosmology. It is particularly important to model the ratio
${P^s_2(k)}/{P^s_0(k)}$ for precision cosmology.

 We find that the results for the two simulation methods are
qualitatively very similar (Figure
\ref{fig:HI_quadpole_by_monopole}). The results from the two methods
are indistinguishable at small $k$ $(\le 0.2 \, {\rm Mpc}^{-1})$ where
the ratio has a value $\sim 1$ or larger. The ratio drops at larger
$k$ where the results from the two methods start to differ, and
becomes negative at even larger $k$.  As noted earlier, the negative
quadrupole is an outcome of the FoG elongation along the line of
sight. We see that the same features can be identified in the results
for both the methods, however they typically are shifted to smaller
$k$ for the HP method due to the larger velocity dispersion in
comparison to the HC method. This is particularly noticeable for the
dip which corresponds to the transition to negative values.

Considering the HC method first, we find that linear theory
(eq.~(\ref{eq:ratio})) with a scale dependent bias matches the
simulations for $k < 0.3, 0.4,0.7 \, {\rm Mpc}^{-1}$ at $z=1,2,3$
respectively.  This range increases to $k < 0.9, 0.7,2.0 \, {\rm
Mpc}^{-1}$ if we include a Lorentzian damping profile. The scale
independent bias models do not provide a very good fit even with a
damping profile.  At $z=4$ all the models are in agreement with the
simulations for $k < 0.4 \, {\rm Mpc}^{-1}$, and they deviate for
larger $k$.  At $z=5,6$ the models match the simulations for a very
limited $k$ range $k < 0.15 \, {\rm Mpc}^{-1}$, and the models
underpredict the ratio for larger $k$. For the HP method,
we find that the linear theory agrees with the simulations for $k < 0.2,
0.3 \, {\rm Mpc}^{-1}$ at $z=1,2$ respectively. This range, however,
increases to $k < 0.6, 1 \, {\rm Mpc}^{-1}$ with the inclusion of
Lorentzian damping. At $z=3,4$ linear theory matches the simulations
for $k < 0.5, 0.4 \, {\rm Mpc}^{-1}$ respectively. This $k$ range does
not improve with the inclusion of any damping profile. The scale
independent bias models fail to provide a good fit to the simulated
ratio at these redshifts even with a damping profile. At $z=5,6$ all
the models match the simulations for a very limited $k$ range $k <
0.15 \, {\rm Mpc}^{-1}$, and the models underpredict the ratio for $k
\lesssim 1 \, {\rm Mpc}^{-1}$.

In summary, we find that models with a scale dependent bias and a
Lorentzian damping profile provide a good fit to the simulations for
large length-scales at $1 \le z \le 4$.  We also find that the complex
nature of the bias is not important over the $k$ range considered
here, and it suffices to consider a real bias with $r=1$.  While our
models work reasonably well for ${P^s_0(k)}$ over the entire $z$
range, the models underpredict ${P^s_2(k)}$ for a considerable part of
the $k$ range at $z \ge 4$.  We note that the bias increases rapidly
with $z$ (Paper I) resulting in a small value of $\beta$ at high
$z$. The discrepancy in ${P^s_2(k)}$ is possibly telling us that the
simple Kaiser enhancement term (eq.~(\ref{eq:modified_kaiser_model}))
which depends only on $\beta$ underpredicts the redshift space
anisotropy at high $z$. The required modification cannot be modelled
through a damping profile which reduces the quadrupole instead of
enhancing it. More sophisticated modelling is possibly needed at high
$z$, and we plan to address this in future work.

Simulating the post-reionization \HI 21-cm signal requires relatively
expensive high resolution cosmological simulations
(Section~\ref{sec:simulating-RSD-HI}) which can be run for a limited
cosmological volume. Here we have provided models which can be used to
compute $P^s_{\HI}(\kpe, \kpa)$ in the range $1 \le z \le 6$, and could
possibly be extrapolated to lower redshifts $0 \le z \le 1$. The
models considered here relate $P^s_{\HI}(\kpe, \kpa)$ to $P(k)$ which
is the dark matter power spectrum. The latter can be computed using
relatively inexpensive low resolution simulations covering larger
cosmological volumes. Our models rely on a scale dependent bias and a
Lorentzian damping profile.  Paper I provides fitting formulas which
can be used to calculate the scale dependent bias, and
eq.~(\ref{eq:sigma_z}) of this paper presents a fitting formula which
can be used to calculate $\sigma_p$ which is used in the Lorentzian
profile. These can be used in eq.~(\ref{eq:modified_kaiser_model}) to
calculate $P^s_{\HI}(\kpe, \kpa)$.

%------------------------------ Bibliography---------------------------------
%\bibliography{reference}

%--------------------- APPENDIX ---------------------------------------------
\appendix{}
\section{}
\label{sec:Appendix-A}
In real space (as against redshift space)  we assume that
$\Delta_{\HI}(\k)$  which refers to the Fourier 
components of the \HI density contrast is related to its matter
counterpart $\Delta(\k)$ through a linear bias parameter   
$\tilde{b}(k)$ whereby 
%---------------------------------------------------------
\begin{equation}
\Delta_{\HI}(\k)=\tilde{b}(k) \Delta(\k) \,.
\label{eq:bias_def}
\end{equation}
%---------------------------------------------------------
Note that the value of $\tilde{b}(k)$ may vary with $k$ (scale
dependent bias) and is, in general, complex (Paper I). We then have 
%---------------------------------------------------------
\begin{equation}
P_{\HI}(k)=b^2(k) P(k)
\label{eq:biad_mod}
\end{equation}
%---------------------------------------------------------
for the real space power spectra where $b(k)$ is the modulus of 
 $\tilde{b}(k)$. Considering the \HI-matter cross-power spectrum
$P_c(k)$ (Paper I) we have 
%---------------------------------------------------------
\begin{equation}
P_c(k)=b_r(k) P(k)
\label{eq:bias_real}
\end{equation}
%---------------------------------------------------------
where $b_r(k)$ is the real part  of   $\tilde{b}(k)$. We also have the
stochasticity defined as $r(k)=b_r(k)/b(k)$. 

In the linear theory of density perturbations, the Fourier  components
of the \HI density contrast in redshift space  is given by 
%---------------------------------------------------------
\begin{equation}
\Delta^s_{\HI}(\k) = \Delta_{\HI}(\k) + f \mu^2 \Delta(\k)
\label{eq:density_fourier}
\end{equation}
%---------------------------------------------------------
where the second term in the R.H.S. incorporates the effect of
peculiar velocities  \citep{kaiser87}.
Using this to calculate the redshift space \HI power spectrum, we have 
\begin{equation}
 P^s_{\HI}(\k)=P_{\HI}(k) + 2 f \mu^2 P_c(k) + f^2 \mu^4 P(k)\,.
\label{eq:rsdps1}
\end{equation}
This can be re-written in terms of the bias and the stochasticity as
 \begin{equation}
 P^s_{\HI}(\k)=(b^2 + 2r b f \mu^2 + f^2 \mu^4) P(k)
 \label{eq:rsdps2}
\end{equation}
which leads to the modified Kaiser term in
eq.~(\ref{eq:modified_kaiser_model}). 
%------- END DOCUMENT
\end{document}